\documentclass{glam-science-paper}

\GLAMTitle{A Private IPFS Data Sanctuary for Verifiable Digital Collection Objects}
\GLAMSubtitle{Recovery, content-derived identity, redundancy, and federated GLAM storage in a seven-node museum prototype}
\GLAMRunningTitle{Private IPFS data sanctuary for digital collection objects}
\GLAMVersion{v0.10}
\GLAMStatus{SCIENTIFIC PREPRINT}
\GLAMDate{September 2026}
\GLAMAuthors{Joachim Snellings, Jasper Funk-Smit, Margot Belot, Eran Wolff, Bernhard Schurian, Sebastian Baldauf, Lisa Pfeiffer, Frederik Berger}
\GLAMAffiliations{ELIO -- Experimental Laboratory for Innovation in Object Digitalization, Museum f\"ur Naturkunde Berlin (MfN)}
\GLAMCorrespondence{joachim.snellings@mfn.berlin}
\GLAMKeywords{GLAM; IPFS; museum digitization; content addressing; distributed file system; raw acquisition; data resilience; deduplication; provenance}
\GLAMRepository{<public release repository>}
\GLAMDataIdentifier{privacy-safe content-operation and four-node telemetry tables}
\GLAMCodeIdentifier{analysis and publication sources in the release package}
\GLAMReproducibility{Kubo 0.36.0; IPFS Cluster 1.1.4; CRDT consensus; claim boundaries and privacy controls documented in the manifest}

\AbstractBackground{Galleries, libraries, archives and museums (GLAM) increasingly produce scientific collection data whose digital state must remain identifiable, recoverable, and verifiable after acquisition. A staging architecture should provide redundancy without requiring every measurement to depend on one workstation or storage path.}
\AbstractMethods{We evaluated a private seven-node InterPlanetary File System (IPFS) deployment at the Museum f\"ur Naturkunde Berlin. Five Raspberry Pi 5 edge peers and two archive-class peers contributed NVMe-backed repositories across two wired zones linked by a shared wireless path. Two real raw-acquisition folders were recovered by content identifier (CID), a large machine-learning model-weight folder was ingested, and sustained telemetry from two edge and two archive nodes was compared.}
\AbstractResults{Raw acquisition A (6.20 GB) materialized in 34.0 s (182 MB/s effective logical rate) and raw acquisition B (5.19 GB) in 14.092 s (368 MB/s). A 156.1 GB model-weight folder was recursively added in approximately 300 s (520 MB/s logical ingest). Four sustained telemetry series comprised 857.3 logged node-hours and 27,616 valid block-stat probes, all reporting successful local block access.}
\AbstractConclusion{The prototype demonstrates a practical institution-local distributed file system (DFS) for content-addressed acquisition staging. Museum-specific object identifiers can remain stable at the collection-object level while each imported measurement automatically receives a content-derived state identity. Replication intentionally distributes identical blocks across failure domains, while content addressing enables reuse of unchanged blocks within repositories. This combination supports verifiable digital collection objects and a federated storage model for collaborating GLAM institutions.}

\begin{document}
\MakeGLAMFrontMatter
\StartGLAMBody

\section{Introduction}

Natural-history digitization produces raw image sets, computed tomography volumes,
label images, meshes, machine-learning artefacts, and other scientific derivatives
that must remain connected to the curated physical object from which they originate
\citep{hedrick2020digitization,belot2026elie}. Before preservation or publication,
these data often pass through acquisition, quality control, reconstruction, and
temporary staging. During that interval, dependence on one workstation or storage path
creates a recoverability problem even when the physical specimen remains safely curated.

We investigate a private InterPlanetary File System (IPFS) cluster as a local
\emph{data sanctuary}: an institution-controlled staging layer that combines
content-derived identity, replicated availability, and restricted peer participation.
IPFS addresses imported blocks through Content Identifiers (CIDs) rather than only
through mutable server paths \citep{benet2014ipfs,daniel2022ipfsfriends}. Public IPFS
studies show that retrieval behavior depends on provider distribution, routing, and
workload \citep{trautwein2022design,shi2024closer}; the present work therefore evaluates
the museum prototype in its own topology instead of claiming a universal IPFS transfer
rate.

Museum collection workflows already assign a museum-specific unique identifier (UID)
to the physical collection object. At MfN, the NURI workflow is one example of
object-level identification \citep{uebeler2022nuri}. The raw measurement folder itself,
however, need not receive an additional manually assigned measurement identifier.
Content-addressed import automatically gives each imported acquisition a root CID,
allowing repeated measurements to remain distinguishable as separate digital states.

\begin{GLAMResearchQuestions}
\textbf{RQ1.} Can a heterogeneous private DFS provide a practical staging and recovery
layer for real museum acquisition data?

\textbf{RQ2.} Can content-derived identity make failed, repeated, accepted, and derived
measurement states verifiable without replacing the museum's established object
identifier?

\textbf{RQ3.} Can the same content identity support deliberate redundancy across
independent storage zones while also enabling block reuse when content is unchanged?
\end{GLAMResearchQuestions}

\section{Methods}

\subsection{Seven-node prototype}

The prototype contained seven physical nodes. Five Raspberry Pi 5 edge systems each
provided 16 GB random-access memory (RAM), 2 TB Non-Volatile Memory Express (NVMe)
storage, and Gigabit Ethernet; Raspberry Pi 5 also provides dual-band Wi-Fi 5
\citep{raspberrypi2026pi5}. Two archive-class systems provided at least eight central
processing unit (CPU) cores, at least 32 GB RAM, and 2 TB NVMe storage.

Approximately 1.0--1.5 TB per peer was configured for the IPFS repository. The nominal
configured raw DFS capacity was therefore approximately 7--10.5 TB before replica
overhead and operational reserve. Table~\ref{tab:hardware} summarizes the configuration.

\subsection{Topology, private swarm, and telemetry}

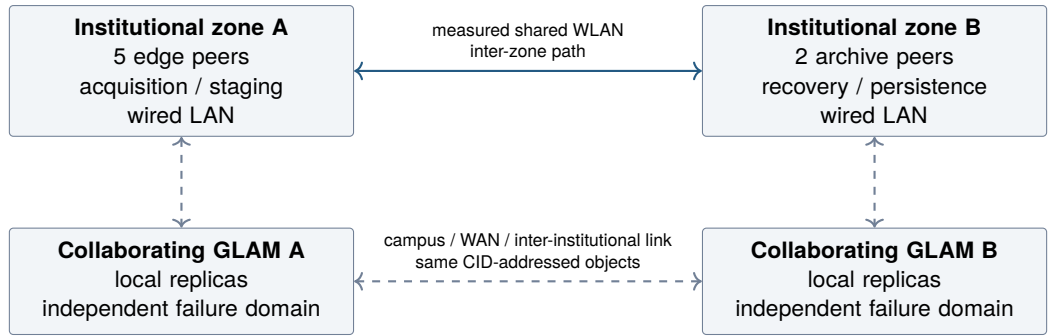
\begin{figure*}[t]
\centering
\begin{tikzpicture}[
  box/.style={draw=GLAMRule,rounded corners=1.5pt,fill=GLAMSoft,
              align=center,inner sep=5pt,font=\sffamily\small,text width=42mm},
  link/.style={<->,thick,draw=GLAMBlue},
  conceptual/.style={<->,thick,dashed,draw=GLAMRule}
]
\node[box] (a) {\textbf{Institutional zone A}\\5 edge peers\\acquisition / staging\\wired LAN};
\node[box, right=46mm of a] (b) {\textbf{Institutional zone B}\\2 archive peers\\recovery / persistence\\wired LAN};
\draw[link] (a) -- node[above,font=\sffamily\scriptsize,align=center]
  {measured shared WLAN\\inter-zone path} (b);
\node[box, below=12mm of a] (c) {\textbf{Collaborating GLAM A}\\local replicas\\independent failure domain};
\node[box, below=12mm of b] (d) {\textbf{Collaborating GLAM B}\\local replicas\\independent failure domain};
\draw[conceptual] (c) -- node[above,font=\sffamily\scriptsize,align=center]
  {campus / WAN / inter-institutional link\\same CID-addressed objects} (d);
\draw[conceptual] (a) -- (c);
\draw[conceptual] (b) -- (d);
\end{tikzpicture}
\caption{Measured two-zone topology and conceptual federation across independent GLAM
failure domains. The lower layer is an architectural generalization, not a WAN
performance measurement. Content identity remains independent of which participating
zone currently serves the object.}
\label{fig:federation}
\end{figure*}

The five edge peers shared one wired local area network (LAN) segment through
\texttt{eth0}; the two archive peers occupied a second wired segment. Cross-zone
communication used a shared wireless local area network (WLAN) path. The measured
arrangement and its institutional generalization are shown in
Figure~\ref{fig:federation}.

The deployment record specifies Kubo 0.36.0 and IPFS Cluster 1.1.4 with a
Conflict-free Replicated Data Type (CRDT) consensus layer. Routing used a Distributed
Hash Table (DHT); multicast Domain Name System (mDNS) discovery was enabled locally,
public bootstrap peers were removed, and a private \texttt{swarm.key} restricted IPFS
participation. An IPFS Cluster secret protected Cluster peer communication. These
controls restrict membership and communication but do not provide file-level
encryption at rest.

Each node additionally exposed intentionally published content through a read-only HTTP gateway behind an Nginx reverse proxy, allowing browser-based retrieval without granting IPFS or Cluster administrative access.

Public IPFS deployments additionally expose routing and availability considerations at the DHT layer. Sridhar et al. demonstrated that manipulation of Kademlia-based provider discovery can interfere with content retrieval and proposed corresponding detection and mitigation mechanisms \citep{sridhar2024censorship}. The present deployment operates under a different threat model because peer participation is restricted to a private swarm.

\begin{table}[!t]
\caption{Seven-node hardware and storage configuration.}
\label{tab:hardware}
\centering\small
\begin{tabularx}{\columnwidth}{@{}lcccY@{}}
\toprule
Class & $n$ & RAM & NVMe & Primary role \\
\midrule
Edge & 5 & 16 GB & 2 TB & acquisition / staging \\
Archive & 2 & $\geq$32 GB & 2 TB & archive / recovery \\
\bottomrule
\end{tabularx}
\end{table}

For compact comparative reporting, sustained stage/probe telemetry from two edge peers
and both archive peers was retained. Edge-01 represents the favorable-WLAN edge case;
Edge-04 provides a second long-duration edge series. Node-state logs report observed
swarm connections and repository state. A block-stat probe queries a known block in
the local IPFS blockstore and therefore measures local content access rather than
peer-to-peer network latency.

\subsection{Scientific workloads and evidence boundary}

The experiment contained three application-level content operations. Recovery A and
Recovery B were real \emph{raw acquisition folders} from museum digitization workflows,
rather than synthetic payloads. The third workload was a substantially larger folder
containing machine-learning model weights.

For a logical object size $S$ materialized in wall time $t$, we report
\begin{equation}
R_{\mathrm{logical}}=\frac{S}{t}.
\label{eq:logical}
\end{equation}
Target blockstore occupancy, provider count, and synchronized interface byte counters
were not retained for the historical retrievals. Equation~\ref{eq:logical} is therefore
an effective logical/materialization rate, not physical Ethernet or WLAN throughput.

Publication derivatives replace operational hostnames, addresses, PeerIDs, and raw
CIDs with generic labels. Private swarm/Cluster secrets and credentials are excluded.

\section{Results}

\subsection{Recovery and ingestion}

\textbf{Raw acquisition recovery A.}
The first recovery workload was a 6.20 GB raw acquisition folder. Materialization on
another participating peer completed in 34.0 s, corresponding to approximately
182 MB/s effective logical rate. The operational significance is that the acquisition
remained addressable through its CID without depending on the original staging path.

\textbf{Raw acquisition recovery B.}
The second raw acquisition folder contained 5.19 GB and materialized in 14.092 s,
corresponding to approximately 368 MB/s effective logical rate. Together, the two
recovery observations show that a repeated or corrected acquisition can coexist with
the previous measurement as a separately identifiable state.

\textbf{Model-weight ingestion.}
A 156.1 GB folder containing machine-learning model weights was recursively added in
approximately 300 s, corresponding to roughly 520 MB/s logical ingest.
Table~\ref{tab:operations} summarizes the three observations.

\begin{table}[!t]
\caption{Historical content operations. Values are single wall-clock observations;
rates are logical/materialization rates, not wire throughput.}
\label{tab:operations}
\centering\small
\begin{tabularx}{\columnwidth}{@{}Yrr@{}}
\toprule
Workload & Size & Observation \\
\midrule
Raw acquisition A & 6.20 GB & 34.0 s; 182 MB/s \\
Raw acquisition B & 5.19 GB & 14.092 s; 368 MB/s \\
Model weights & 156.1 GB & $\sim$300 s; $\sim$520 MB/s \\
\bottomrule
\end{tabularx}
\end{table}

\subsection{Sustained four-node telemetry}

The four comparative telemetry series provide 857.3 combined logged node-hours,
10,248 stage samples, and 27,616 valid block-stat probes. All retained valid
block-stat probes reported successful local block access. Because these windows were
not synchronized, the sum of node-hours is descriptive coverage rather than cluster
uptime.

Table~\ref{tab:telemetry} shows the compact comparison. Archive-node block-stat
latencies were lower and tighter in these logs than the two edge-node series, while
Edge-01 showed a broader P90 tail despite a similar median to Edge-04.

\begin{table}[!h]
\caption{Sustained node-level telemetry used for compact comparative reporting.
P50/P90 are block-stat latencies, not network ping latency.}
\label{tab:telemetry}
\centering\small
\begin{tabularx}{\columnwidth}{@{}YrrrY@{}}
\toprule
Node & Window h & Stage $n$ & Probes & P50/P90 ms \\
\midrule
Edge-01 & 312.19 & 3,736 & 9,084 & 28.2 / 87.8 \\
Edge-04 & 313.50 & 3,739 & 10,965 & 30.0 / 35.6 \\
Archive-01 & 67.10 & 805 & 1,881 & 15.5 / 18.5 \\
Archive-02 & 164.49 & 1,968 & 5,686 & 14.3 / 14.9 \\
\bottomrule
\end{tabularx}
\end{table}

\begin{GLAMEvidence}
RQ1 is supported at the application layer by two raw-acquisition materialization
observations, one large model-weight ingest, and sustained blockstore telemetry on both
edge and archive hardware. The experiment demonstrates practical operation of the
private DFS but does not establish physical WLAN saturation or a transfer-rate
distribution.
\end{GLAMEvidence}

\section{Discussion}

\subsection{From museum object to verifiable digital measurement state}

The physical collection object continues to carry its museum-specific UID. Content
addressing adds a second, automatically generated identity at the measurement level:
the imported acquisition folder receives a root CID representing that exact imported
state.

Figure~\ref{fig:lineage} illustrates the consequence for repeat acquisition. If the
first measurement is incomplete or scientifically unsuitable, it can remain retained
as acquisition A. A repeat becomes acquisition B with a distinct CID, and downstream
reconstruction can explicitly reference the accepted acquisition rather than a mutable
folder name such as ``final'' or ``scan2''.

\Needspace{19\baselineskip}
\begin{center}
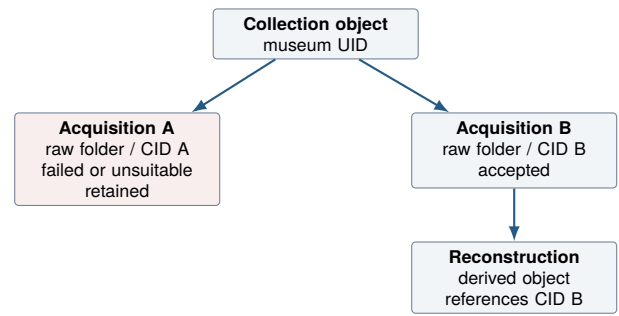

\begin{minipage}{0.98\columnwidth}
\centering
\begin{tikzpicture}[
  box/.style={draw=GLAMRule,rounded corners=1.5pt,fill=GLAMSoft,
              align=center,inner sep=3pt,font=\sffamily\scriptsize,text width=25mm},
  fail/.style={draw=GLAMRule,rounded corners=1.5pt,fill=GLAMRose,
              align=center,inner sep=3pt,font=\sffamily\scriptsize,text width=25mm},
  arrow/.style={-{Latex[length=2mm]},thick,draw=GLAMBlue}
]
\node[box] (uid) {\textbf{Collection object}\\museum UID};
\node[fail, below left=7mm and -1mm of uid] (a) {\textbf{Acquisition A}\\raw folder / CID A\\failed or unsuitable\\retained};
\node[box, below right=7mm and -1mm of uid] (b) {\textbf{Acquisition B}\\raw folder / CID B\\accepted};
\node[box, below=7mm of b] (d) {\textbf{Reconstruction}\\derived object\\references CID B};
\draw[arrow] (uid) -- (a);
\draw[arrow] (uid) -- (b);
\draw[arrow] (b) -- (d);
\end{tikzpicture}
\captionof{figure}{Repeat-acquisition lineage. The museum UID identifies the physical
collection object; content addressing gives each imported measurement folder its own
digital state identity. A failed acquisition can remain preserved without becoming the
source for reconstruction.}
\label{fig:lineage}
\end{minipage}
\end{center}

A folder CID is not merely a conventional file hash: the UnixFS root also reflects the
directory representation and can depend on import/chunking parameters. For preservation,
a compact manifest should therefore retain the root CID, relevant import parameters,
and an independent fixity digest such as SHA-256.

\subsection{Redundancy and deduplication serve different purposes}

Replication and deduplication are complementary rather than contradictory. Replication
deliberately stores identical content in independent failure domains so that the loss
of one node or zone does not remove the only available copy. Content-addressed block
reuse avoids storing an additional unique block when the same block already exists in
a repository. IPFS therefore provides block-level deduplication for identical content,
while Cluster replication intentionally creates additional physical copies where
resilience requires them \citep{benet2014ipfs,shi2025centralization}.

For related versions of an acquisition or derivative, unchanged blocks can be reused
when their block identities remain unchanged. The amount of reuse is workload- and
chunking-dependent; this experiment did not measure a deduplication ratio. Highly
compressed or substantially rewritten files may share relatively few blocks.

\subsection{Federated GLAM storage}

RQ3 extends the measured two-zone topology into an institutional collaboration model.
Because a CID identifies content rather than a particular server path, collaborating
GLAM institutions can retain replicas of agreed digital collection objects while
referring to the same content identity. The resulting shared content-addressed
collection-data layer can be geographically or administratively zone-independent.

This does not require every institution to expose its complete internal repository.
Each participant can retain local acquisition replicas and exchange only agreed
content, manifests, or derivatives. Storage responsibility can therefore be distributed
incrementally across participating institutions and commodity hardware instead of
requiring one central staging appliance. More broadly, distributed edge–cloud architectures require explicit decisions about which processing and data-management operations remain local and which are delegated across the infrastructure. ContinuumConductor formalizes such decentralization decisions for heterogeneous edge–cloud environments according to resource, responsiveness, and privacy considerations \citep{reiter2025continuum}. Although developed for process mining rather than scientific collection storage, this systems perspective is relevant to future federated GLAM infrastructures.
The prototype motivates this potentially cost-distributable model, but no total-cost-of-ownership comparison was performed.

A slower campus or wide area network (WAN) link affects recovery time, not the
identity of the data being recovered. If one institutional zone loses its local copy,
another zone retaining the required pinned blocks can reseed replacement peers after
connectivity is restored. A CID alone does not guarantee persistence; replica placement,
pinning policy, credentials, and independent backup remain operational requirements.

\begin{GLAMInterpretation}
RQ2 is answered at the identity level: content addressing converts each imported
measurement into an immutable-by-reference state. Changing the represented acquisition
produces a different content identity rather than silently changing the referenced
state. RQ3 is supported architecturally by location-independent CIDs and deliberate
replication, but inter-institutional recovery was not benchmarked in this experiment.
\end{GLAMInterpretation}

A CID alone does not guarantee persistence; replica placement,
pinning policy, credentials, and independent backup remain
operational requirements.

More generally, heterogeneous distributed storage systems must account for differences in capacity, performance, and failure characteristics between participating resources. D-Rex demonstrates a reliability-aware approach in which data placement and erasure-coding parameters are adapted to heterogeneous storage nodes \citep{gonthier2025drex}.

\subsection{Capacity and network scaling}

Distributed storage systems must account not only for aggregate capacity but also for heterogeneity in node performance, storage capacity, and failure characteristics. Recent work such as D-Rex explicitly incorporates these differences into reliability-aware data placement and erasure-coding strategies for heterogeneous distributed storage resources \citep{gonthier2025drex}.

With approximately 1.0--1.5 TB configured per peer, the seven-node prototype provides
roughly 7--10.5 TB nominal raw capacity. Under balanced replication factor $r$, unique
replicated content is approximately bounded by
\[
C_{\mathrm{unique}} \lesssim \frac{C_{\mathrm{raw}}}{r},
\]
before reserve and any deduplication benefit.

Network bandwidth scales differently. Several peers can contribute concurrent flows
over a switched LAN until a shared network interface controller (NIC), switch fabric,
uplink, storage device, or CPU becomes the bottleneck. A 10 GbE switch fabric or
stacked 10 GbE uplink can raise aggregate LAN capacity, while an individual 1 GbE
endpoint remains limited by its own interface. WLAN is less additive because peers
share airtime and access-point capacity. The historical experiment did not independently
measure the WLAN ceiling.

At larger peer populations, data availability additionally depends on network conditions, peer responsiveness, and dissemination strategy.
PANDAS demonstrates a peer-to-peer availability protocol designed to tolerate message loss, node failures, and unresponsive participants while scaling to large peer populations \citep{pigaglio2025pandas}. 
Although its application domain differs from persistent scientific storage, these systems-level constraints are relevant to future multi-institutional federation.

\section{Limitations}

\begin{GLAMLimitations}
\begin{small}
The two raw-acquisition recovery values and the model-weight ingest value are single
historical observations. Target blockstore state, provider count, exact
source/destination zone, and interface byte counters were not synchronized to those
runs. The logical rates therefore cannot be interpreted as physical network
throughput.

The four telemetry windows are sustained but asynchronous. Block-stat probes measure
local blockstore access rather than P2P retrieval. Deduplication is an architectural
property discussed from the content-addressed storage model; no block-reuse ratio was
measured. Cross-institutional federation and full-zone restoration are design
extensions rather than experimentally demonstrated outcomes.
\end{small}
\end{GLAMLimitations}

\section{Conclusion}

This work evaluates a private seven-node IPFS data sanctuary for scientific
digitization in a GLAM institution. Two real raw-acquisition folders were recovered by
CID at multi-gigabyte scale, a 156.1 GB model-weight folder was recursively ingested,
and sustained telemetry from two edge and two archive nodes documented prolonged
operation of the local content-addressed layer.

The central contribution is the combination of verifiable digital measurement identity
and distributed custody. The museum object retains its established UID, while each
imported measurement acquires a separate content-derived CID. Failed, repeated,
accepted, and derived states can therefore remain distinguishable without overwriting
the scientific record. Replication places selected states across failure domains for
resilience, while block-level content addressing enables deduplication when identical
blocks recur.

The same mechanism can support a federated GLAM data sanctuary in which participating
institutions contribute storage capacity and recovery responsibility without changing
the identity of the shared digital collection objects. This makes the storage layer a
candidate substrate for affordable, distributed, and verifiable scientific collection
data management, while leaving cost and inter-institutional recovery performance for
future evaluation.

\section*{Abbreviations}
{\footnotesize
\begin{tabularx}{\columnwidth}{@{}lY@{}}
\toprule
Term & Full form \\
\midrule
CID & Content Identifier \\
CPU & Central Processing Unit \\
CRDT & Conflict-free Replicated Data Type \\
DFS & Distributed File System \\
DHT & Distributed Hash Table \\
GLAM & Galleries, Libraries, Archives and Museums \\
IPFS & InterPlanetary File System \\
LAN & Local Area Network \\
mDNS & multicast Domain Name System \\
NIC & Network Interface Controller \\
NURI & MfN URI used in the museum object-identification workflow \\
NVMe & Non-Volatile Memory Express \\
P2P & peer-to-peer \\
RAM & random-access memory \\
UID & unique identifier \\
WAN & Wide Area Network \\
WLAN & Wireless Local Area Network \\
\bottomrule
\end{tabularx}
}

\section*{Data, privacy, and reproducibility}
\PrintGLAMArtifactRecord

Operational IP addresses, hostnames, raw PeerIDs, private swarm keys, Cluster secrets,
credentials, and unreleased CIDs are excluded from the publication package. The
institution-internal raw logs remain under local control.

\section*{Acknowledgements}
We thank colleagues in ELIO, institutional IT, collections, and digitization teams.
We gratefully acknowledge European Union funding of the ELIO project and the
infrastructure provided by the Museum f\"ur Naturkunde Berlin.

\balance\small
\bibliographystyle{plainnat}
\bibliography{references}

\end{document}